# Mineralogy of a Sulfate-rich "Inverted Channel" in the Atacama Desert, Chile: Clues to its Formation and Preservation

E. Z. Noe Dobrea[1], R. M. E. Williams[1], W. E. Dietrich[2], A. D. Howard[1], J. C. Cawley[3], and R. P. Irwin III[3]

[1]Planetary Science Institute, 1700 E. Fort Lowell, Suite 106, Tucson, AZ 85719
[2]Earth & Planetary Science, University of California—Berkeley, 307 McCone Hall, Berkeley, CA 94720
[3]Center for Earth and Planetary Studies, National Air and Space Museum, Smithsonian Institution, Independence Ave at 6th St, SW, Washington, DC 20560

**Abstract**
We have performed a stratigraphic and mineralogical analysis of a vertical transect across a ridge located at the distal end of a system of eroded alluvial deposits in the northern Atacama Desert of Chile.  The ridge, which is interpreted to be an inverted channel, exhibits a history of sedimentary, evaporitic, and diagenetic origin that includes groundwater mobilization and precipitation of anhydrite cements throughout the volume of the ridge.  The ridge consists of two units: a lower one exhibiting a sedimentary and diagenetic history, and an upper one exhibiting an evaporitic history.  Interbedded in the section are also anhydritic and gypsic paleosols. Two mechanisms that contribute to channel preservation and inversion are identified in this case. The first mechanism is the cementation of the volume by anhydrite cements during early diagenesis, and the second newly identified mechanism is the armoring of the lateral slopes of the ridge by halite-rich cement.  The slope-conforming armor formed by this second mechanism developed subsequent to the formation of the ridge as a consequence of the remobilization of soluble salts.  Finally, we identify a series of Ca-sulfate-rich plates on the surface of the ridge, which we interpret here to form by fracturing and subsequent erosion of an evaporitic deposit. The plates exhibit a reticulated surface texture, which we interpret as the result of periodic deliquescence and reprecipitation of a thin surface film of the evaporite deposits in response to thick morning fogs that occur in this part of the Atacama.  The cross section of the plates exhibits a thin portion of biological material, which we ascribe to bacterial mats that take advantage of the deliquescence of the substrate to obtain their water.  This later has important implications in the search for extant or extinct life on Mars.

**Introduction**
Inverted channels are a geomorphic feature that results from differential erosion, occurring when the cemented or armored floor of a fluvial channel erodes at a slower rate than the surrounding, more erodible terrain (Zaki *et al.,* 2018).  Over time, these negative-relief features evolve into positive relief.  Although they present geomorphologically as ridges, they can often be identified in plan view by their sinuosity or by branching patterns that mimic those of fluvial features.  The development of inverted channels with weakly cemented or armored floors is most effective in arid environments dominated by aeolian deflation.  Given their preservation of morphometric character and sediments, inverted channels provide the opportunity to investigate past climate, hydrology, and aqueous chemistry (Pain and Ollier, 1995).

Inverted channels are regularly observed in regions of Mars where fluvial systems were present and have since been eroded by wind (e.g., Morgan *et al.*, 2014; Davis *et al.*, 2018; Di Pietro *et al.,* 2018). Hence, the study of the distribution and morphology of sinuous ridges on Mars has been used to constrain the timing and nature of fluvial activity on Mars (*e.g.,* Malin and Edgett, 2003; Williams, 2007; Bull *et al.*, 2010). Understanding the mechanisms by which channels are preserved helps us to better constrain the environmental and aqueous conditions present during the time of channel formation on Mars.

Here, we describe the mineralogy and stratigraphy of the WIN ridge, an inverted channel located in the Pampa de Tamarugal, Atacama Desert, Chile. We document the the stratigraphy including bedding style and thickness, mineralogy, and degree of induration of the exposed layers and present a model for the geological history of this ridge. We also show cementing agents throughout its volume and the formation of slope-conformal salt armor along its escarpments have contributed to the preservation of the channel bed as a ridge.

**Setting:**
The Pampa del Tamarugal is a geologic province located within the Central Depression of the Atacama Desert, between 19°30' and 22°15' south latitude and bounded by the low (<2 km) Coastal Range to the west and the towering Andes Mountains (>3.5 km) to the east (**Fig. 1a**). These mountain ranges are barriers to atmospheric moisture reaching the elevated plateau of the Central Depression, such that the Atacama has been one of the driest places on Earth for at least the last 13 Myr (Clarke, 2006). Presently, the general environmental characteristics of the region are the hyperaridity, strong winds, and exceptionally little rainfall, which occurs as sporadic torrential rains. Additional sources of water in the region are the thick coastal fogs known as the Camanchacas, which can penetrate up to 100 km inland, and water sourced from occasional winter storms in the Andes to the east.

With little direct precipitation in the Atacama Desert (<3 mm/yr; e.g., Clarke, 2006; Rech *et al.,* 2006), both surface water and groundwater are sourced from high in the Andes Mountains (*e.g.,* Houston, 2001, 2002, 2009; Houston and Hart, 2004). The timescales for Andes storm water to reach the interior Atacama Desert vary by route, with groundwater emerging to within 5 meters of the surface along deep fracture sets in $10^4$–$10^5$ years, and more frequent surface runoff events occurring on timescales of $10^0$–$10^1$ years (Jayne *et al.*, 2017). Streams form from precipitation in the Andes Mountains and transport sediment approximately 150 km westward into the basin to form large alluvial fan complexes (Morgan *et al.*, 2014).

Infilling within the Pampa del Tamarugal basin is a sequence of alternating volcanic and nonmarine sediments comprising Altos de Pica, El Diablo, Quillagua, and Soledad formations, with deposits from the Late Oligocene to Pleistocene (Dingman and Galli, 1965; Houston, 2002; Magaritz *et al.,* 1989; Nester, 2008; Rojas and Dassargues, 2007; Quezada *et al.,* 2012). Sedimentary deposits near the terminal basin Salar de Llamara show relatively young cycles of deposition and erosion. 'Ancient' alluvial and playa deposits were emplaced during periods of enhanced precipitation in the Andes Mountains during the Early Holocene and Late Pleistocene, and more recent 'active' alluvial deposits cut the older, abandoned fan surfaces (Quezada *et al.,* 2012; Nester *et al.,* 2007; Santoro *et al.,* 2017). Wind dissection of playa deposits has sculpted furrows and yardang fields, and it has further expanded dissolution pits (Williams *et al.,* 2016).

Short (<2 km) ridge segments on the distal bajada are interpreted as the wind-eroded remnants of fan distributary channels ('inverted channels') because the scale and configuration is consistent with nearby modern incised channels on the alluvial fan (**Figs. 1b,c;** Morgan *et al.,* 2014). We examine the mineralogical and sedimentological record for one ridge segment, WIN (21.119°S, 69.578°W), and a nearby yardang (21.1171°S, 69.5769°W).

**Methods:**
Excavation of the vertical transect at WIN ridge was performed using hand shovels and rock hammers. Measurements of layer thickness used a Trimble R8 differential global positioning system (GPS) base station and rover, and a Trimble GeoXT handheld GPS receiver capable of sub-meter post-processed accuracy and in-field geographic information system (GIS) capability. Tape measurements were also used to track sampling locations. The texture, color, and degree of induration were observed and recorded. *In-situ* spectral measurements were obtained using an ASD FieldSpec 3 with a contact probe. Collected samples were later measured in the lab using the same spectrometer as well as with an Olympus Terra portable X-Ray diffractometer (CuKα; Blake *et al.*, 2012). X-Ray diffraction (XRD) analysis was performed using the commercially available software XPowder. Quantitative estimates of relative abundances derived from the XRD pattern were performed using Reference Intensity Ratios methods. Twenty one specimens from WIN ridge and 6 specimens from the nearby yardang were analyzed by XRD. Samples were selected to capture mineralogy changes within the stratigraphic section.

Spectral measurements were acquired in the visible through near-infrared (VNIR – 350 – 2500 nm) wavelengths, converted to reflectance relative to Spectralon, and analyzed by using a combination of Tetracorder and visual comparison of the calibrated spectra with laboratory spectra available in the U.S. Geological Survey spectral library (Kokaly *et al.*, 2017). Tetracorder is an expert system capable of performing mineralogical interpretation and mapping of spectral data. The system compares a given spectrum against spectra from a spectral library, fitting for multiple bands and generating robust identifications. The Tetracorder algorithms and expert system are described in detail in Clark *et al.* (2003), and the algorithm accuracy was extensively evaluated in Swayze (1997) and Swayze *et al.* (2003). Additions and refinements to Tetracorder are described in Clark *et al.* (2015; 2016) and Clark (2019). Tetracorder has been central to dozens of studies on the Earth (Clark *et al.*, 2001; 2006; Clark *et al.*, 2010; Kokaly *et al.*,2003, 2007, 2011 Kokaly *et al.*, 2011; Livo *et al.*, 2007; Swayze *et al.*, 2000, 2014 and references therein), the Moon (Pieters *et al.*, 2009; Kramer *et al.*, 2010; Clark *et al.*, 2009), Mars (e.g., Ehlmann *et al.*, 2016), and icy satellites in the Jupiter and Saturn systems (e.g., Carlson *et al.*, 1996; Clark et al, 2003 Clark *et al.*, 2012).

**Observations and mineralogical mineralogical constraints:**
<u>WIN ridge</u> is approximately 1 km long, about 1.20 m tall, flat-topped, and bounded by lateral slopes of about 20º. It is located within a field of yardangs, most of which are lower in height than the ridge (**Fig. 2**). The terrain between yardangs is flat and fractured, forming decimeter-sized polygons (**Fig. 2D**). **Figure 3** shows the transect that was made on the east side of WIN ridge. **Table 1** lists the layers that we identified based on color, grain size, degree of induration, texture, and mineralogy inferred from XRD and VNIR spectral measurements.

Layers within WIN ridge were observed to exhibit a complex stratigraphy involving variations in grain size, bedding, colors, and mineralogy. Primary minerals, alteration products, and evaporitic salts were detected in discreet horizons using XRD, and spectral observations provided complementary mineralogical information. **Figure 4** plots the relative XRD abundances of the reported minerals and indicates their relative stratigraphic position within the ridge, **Figure 5** compares the VNIR spectra of the different layers to laboratory spectra of inferred minerals, and **Figure 6** shows a simplified stratigraphic section for clarity.

The entire ridge volume was found to be well-indurated with a dominance of sulfate mineralogy. Anhydrite was found to be present throughout the section in concentrations as low as 30 wt% in the lower unit and almost 95 wt% in the upper unit. XRD amorphous materials were consistently found to constitute about 10 wt% of the bulk.

Excavation of the ridge's transect encountered two hardpans, one occurring at the ground level (S1) and a second one at 30 cm above the ground level (S5). These hardpans were challenging to break through, resulting in a stair-step transect. Some of the layers above the hardpan (S5 in **Fig. 3**) exhibited a concave-up shape (**Fig. 7**) (S9 and perhaps S8), although it is unclear whether S6 and S7 exhibit the upward concavity.

The ridge was found to be indurated throughout its volume, and capped by a discontinuous, thin (~1 cm) layer of indurated plates that exhibited a curious reticulation on the exposed side (**Fig. 8**), while their underside transitioned into a soft, loose, light-toned layer of salts. At the edges of the ridge's top, the reticulated texture of the platy material transitioned into a digitate morphology (**Fig. 9**), although the reticulation is retained (**Fig. 9B**). The side slopes of the ridge were thinly mantled with unconsolidated material which, when scraped away, revealed a cm-thick slope-conforming indurated layer of halite (S8b) that armored the slope (**Fig. 10**).

In broad terms, we found that the ridge could be compositionally divided into a lower (S2-S9) and an upper (S10-S15) unit, with the main distinction based on presence or absence of primary minerals. The lower unit is characterized by the presence of primary minerals (quartz, feldspar and mica) and sulfates (anhydrite and jarosite). The upper unit is characterized by sulfates (anhydrite and gypsum) and other evaporitic salts (halite and glauberite) with quartz present in minor abundances. Amorphous materials were found by XRD to be present throughout at abundances of about 10 wt%. Although not detected by XRD, the VNIR spectra of most samples exhibited absorptions consistent with montmorillonite.

*WIN Lower Unit:*
The lower unit consists of 5 distinct fine sand layers (S2, S4, S6, S8, S9) of varying thickness (6 to 30 cm) and varying levels of post-depositional diagenesis. One 25 cm sandy mud layer (S7) separates the three lower sand layers from the two upper ones. The bottom subunit (S2-S4) was sandwiched between two hardpans (S1 and S5), and the top subunit (S6-S9) was sandwiched between hardpan S5 and a horizon composed almost exclusively of anhydrite (S10). The main difference between the top and bottom subunits was that the top subunit exhibited extensive mottling and differences in mineralogy (*e.g.*, the presence of jarosite).

The hardpans S1 and S5 consisted dominantly of about 75 wt% anhydrite and 15 wt% primary minerals. Of these, the upper (S5) hardpan was characterized by the additional presence of halite in minor abundances (5 wt%). Spectroscopically, hardpans S1 and S5, as well as layers S9 and S10, exhibited spectra consistent with montmorillonite on the basis of an Al-OH combination band centered at 2205 nm, an OH stretch overtone centered at 1410 nm, and an $H_2O$ combination band centered at about 1920 nm (**Fig. 5**). We noted that none of the observed bands exhibit the sharpness of the equivalent bands exhibited by the spectral library sample, suggesting that the observed montmorillonite was poorly crystalline. The absence of the 001 XRD peak around 2θ ~ 7°, which results from interlayer spacing of smectites, supports the idea that montmorillonite occurs in an amorphous phase and at low abundances (given that the total concentration of amorphous material was about 10 wt%). Additionally, we also noted that the ~1920 nm combination band was typically centered at wavelengths slightly longer than that of the library montmorillonite. This shift is diagnostic of the presence of other minerals that contribute to the spectrum. Sulfates typically absorb at longer wavelengths (1930 nm and up) than smectites, and it would be no surprise if these sulfates were responsible for the shift. Candidate sulfates include anhydrite (on the basis of a ~1930 band), bassanite (on the basis of the position of the 1921 nm band in S1, S9, and S10 and a possible muted band at 1768 nm), and gypsum (on the basis of a shift of the 1920 nm band to 1940 nm in S5). We note that XRD measurements were consistent with gypsum and anhydrite but did not with any appreciable concentration of bassanite.

Between these two hardpans was the bottom subunit: a thick (40 cm) layer (S2-S4) that exhibited white flakes near the base, decreasing in concentration upwards. This layer was also observed to exhibit cracks that were filled by orange and white rinds (**Fig. 11**), as well as occasional orange-colored stains and small (mm-scale) holes. The surface of hardpan S1 at the contact with S2 exhibited a cm-scale roughness (**Fig. 12**), whereas the contact between the top of S4 and the hardpan S5 exhibited a scalloped surface onto which the hardpan was draped. The scallops had a scale of about 6 cm in diameter (**Fig. 13**). XRD analysis revealed that S2-S4 contained 50–60 wt% anhydrite and about 30–40 wt% primary minerals. VNIR spectra exhibited absorptions consistent with mica (illite or muscovite) centered around 1410, 1910, 2205, and 2350 nm (although contributions by montmorillonite cannot be excluded), as well as pyroxene on the basis of an upward concavity in the 1000–2000 nm region.

A jarosite-bearing horizon (S6-S8) was found directly above S5 hardpan. This horizon consisted of three layers exhibiting different characteristics with jarosite abundance ranging from ~10–45 wt% in XRD. Layer S6 was a dark-toned purple, well-sorted sandstone with red iron mottling and root holes. Layer S7 was a highly heterogeneous sandstone with red, purple, orange, and yellow mottling (**Fig. 14**); and XRD measurements from five samples in this layer showed significant compositional heterogeneity within the layer, and the highest concentration of jarosite associated with the orange material (**Fig. 4**). Layer S8, the orange layer in **Figures 3 and 15**, consisted of Fe-stained sands and clays, exhibited mottling, and was peppered with white flakes. Spectroscopically, only the spectra of layer S8 exhibited absorptions at about 2215 nm and 2265 nm consistent with jarosite. The spectra of S6 and S7 were similar to the spectra of the bottom subunit and were consistent with montmorillonite. Some structure in the long-wavelength wing of the 2210 nm feature was inferred to be due to jarosite.

This jarosite-bearing horizon was in turn overlain by a gypsum-bearing horizon (S9), which capped the bottom unit. Layer S9 consisted of fine sands (with mica), clay, and large Fe-stained root casts. It contained the lowest concentration of anhydrite (<5 wt%) in the stratigraphy and a relatively high concentration of gypsum (about 40 wt%). Spectroscopically, the layer was practically indistinguishable from the hardpan S1.

*WIN Upper Unit:* The upper unit was characterized by an absence of primary minerals in XRD and a preponderance of salts. The base of the upper unit (S10) exhibited a soft, friable, granular material containing almost 95 wt% anhydrite. Overlying this material was a horizon containing glauberite and halite lenses (S11-S13) at concentrations of a few wt% in S11 to about 10 wt% in S12 and S13, with the concentration of glauberite increasing upsection. The remainder of the mineralogy was dominated by anhydrite at abundances of 70 to almost 90 wt%. Layer S11 consisted of very dark (black) sandstone. Layer S12 consisted of lightly cemented, horizontally layered slabs of salt-rich deposits (this layer correlates to a peak in halite abundance in S12 relative to S11 and S13); S13 no longer contained these salt lenses, but instead exhibited reddish Fe-stains. Spectroscopically, we noted a transition from spectra consistent with montmorillonite (exhibiting an Al-OH combination band at 2205 nm) in layer S11 to spectra consistent with the Mg-smectite saponite (exhibiting Mg-OH combination bands at 2312 nm and 2387 nm, as well as a doublet at 1392 nm and 1413 nm) (**Fig. 5**). The saponite bands increased in contrast from S12 to S13, correlating with the increase in the abundance of glauberite. A rounded absorption around 2205 nm was interpreted to be due to the presence of hydrated silica (opal), although it could also be due to small amounts of montmorillonite. We lean toward the hydrated silica interpretation on the basis of spectral fits performed with Tetracorder.

Overlying this glauberite + halite + saponite horizon was a layer of soft, loose salt dominated by gypsum (S14), which was in turn capped by decimeter-scale plates (caprock – Fig. 8a) and other detritus (S15). Whereas Ca-sulfates with varying degrees of hydration were identified in this capping material, the decimeter-scale plates also contained easily detectable, although minor amounts of bassanite ($2CaSO_4 \cdot H_2O$). This detection contrasts with the rest of the materials, which did not contain readily detectable amounts of bassanite. Spectra of these materials (S14 and S15) exhibited multiple features consistent with gypsum.

*Yardang:* We also performed a vertical transect of a nearby yardang to compare to the bedding and mineralogical stratigraphy of the WIN ridge. The yardang deposits showed much less post-depositional alteration and thus likely record original primary mineralogy of sediment in this area (**Fig. 16**). The yardang was located about 14 meters northwest from our sampling location at the WIN ridge. The yardang was about 82 cm tall, with the lower half consisting of two (~15 cm thick) layers dominantly of well-sorted, fine-grained sand (Y1-Y2), and an upper half of two (~25 cm thick) dominantly clay- and silt-sized grains with some sand (Y3-Y4). Root casts and orange stains were observed in the upper half of the yardang. We sampled the four distinct layers and constrained the mineralogy using XRD and VNIR spectroscopy (**Table 2, Fig. 4**). We found that anhydrite and halite were present in the highest abundances near the bottom and decreased in abundance with height, whereas gypsum was absent at the bottom and increased in abundance with height. Quartz and plagioclase were major components of the yardang as well, with muscovite appearing as a minor component throughout. An increase in these primary minerals' concentration with height was observed, in contrast with the WIN ridge, where plagioclase and muscovite exhibited relatively constant concentration up to 75 cm, above which they were not detected.  Compared to the lower WIN ridge unit, the yardang exhibited significantly higher abundances of plagioclase (2x-3x greater), but comparable amounts of muscovite and quartz.

**Discussion:**
The WIN ridge preserves evidence for a complex history of aggradation, fluvial erosion, groundwater diagenesis, and aeolian erosion interspersed with periods of paleosol formation. Multiple mineralogical horizons and paleosols were identified, indicating a protracted history of early-stage diagenesis.

Horizons and paleosols:
As described in the Results section, the WIN ridge can be divided into two sections on the basis of mineralogy: a lower unit containing sulfates and primary minerals, and an upper unit containing sulfates and other evaporites.

*WIN Lower unit:* The primary mineralogy of the lower unit — feldspar, mica, and quartz — is consistent with that of the nearby yardang, and it likely constitutes the original mineralogy before diagenesis.  The observed anhydrite hardpans are interpreted as paleosols formed from the evaporation of standing water.  The presence of halite in the S5 hardpan supports this hypothesis, while the root casts observed in the rocks between hardpans S1 and S5 support the hypothesis of a wet depositional environment for S2-S4.  Additional root casts and gypsum found above the jarosite-bearing layers (S6-S8) indicate a second period of wet conditions and subsequent evaporation (S9).  Why this evaporative period formed a gypsum-dominated paleosol rather than an anhydrite-dominated one, is not clear.  Possible reasons include a lower salinity of the water during precipitation, or post-depositional transformation of anhydrite to gypsum in humid conditions.

Evidence for diagenesis is observed in the jarosite horizon (S6-S8).  Notable mottling is present in these layers, with strong red and purple colorations due to Fe-stains, as well as yellow patches and white salts (**Fig. 14**).  Evidence for bioturbation is apparent in the sandy mud of S7.  The presence of a jarosite horizon is particularly intriguing, as jarosite commonly forms in acidic environments (pH 2-4 *e.g.*, King and McSween, 2005) from the oxidation of iron sulfides. Hence, its presence in the stratigraphy could imply either a period of acidification of the groundwater as a consequence of an increase in $SO_2$ due to a volcanic eruption in the region, or by the local formation of pyrite during diagenesis of buried organics (volcanoes are very common in the Andes east of the area, and eruptions are likely to have occurred during the period in which the layers were forming around 9 Kya (Williams *et al.*, 2020).  Jarosite would subsequently precipitate in the presence of oxidizing conditions forming yellow to orange patches, and over time it could turn to red as it oxidized to hematite.

Jarosite in paleosols has been previously observed to occur as infillings of channels (Buurman, 1975) and pore spaces (Moorman and Eswaran, 1978).  In these cases, jarosite forms from the oxidation of pyrite during exposure to oxidizing conditions (*e.g.,* emersion of the host sediment). Pyrite itself forms from the reduction of sulfur in an anoxic environment by sulfate-reducing bacteria and in the presence of dissolved iron and gypsum. Here, bacteria reduce sulfur to sulfide, and in the presence of dissolved iron, pyrite is formed. These sulfate-reducing bacteria also need other nutrients to live, which are provided by the buried organics. Hence, in the presence of sulfur- and iron- bearing groundwater, pyrite will precipitate around buried organics as a biologically mediated deposit (*e.g.*, Zhao *et al.*, 2017; Thiel *et al.*, 2019). The groundwater that percolated through the lower unit was clearly rich in dissolved sulfur and likely contained the requisite iron for the formation of sulfide. The roots are testament to the above process, specifically the presence of orange rinds (interpreted to be jarosite) surrounding some of the root tracks in S2 and S9. Hence, although S8 lacks the obvious root tracks, we posit that this layer may have been organic-rich, leading to the mineralization of jarosite as a consequence of its biogenically mediated degradation.

*WIN Upper Unit:* The mineralogy of the upper unit lacks the primary minerals present in the lower unit, and it is considered to represent an aggradational evaporative sequence that formed subsequent to the formation of the fluvial channel. It consists of horizons of evaporitic minerals, saponite, hydrated silica, and minor amounts of quartz. The quartz is likely to have been merely aeolian infill during the aggradation period. The lack of primary minerals implies that there were no surface overland flows involved in the emplacement of the upper unit.

The evaporative sequence consists dominantly of anhydrite, with over 90 wt% anhydrite in the bottom layer and decreasing to about 70 wt% in the layers containing glauberite and halite in minor amounts (up to 15 wt% and 10 wt%, respectively). Both of these salts can form in continental and marine deposits, both are very soluble, and glauberite can readily dissolve and re-precipitate as gypsum in sulfur-bearing waters. In a thorough study of evaporitic sequences in the salars of the Atacama Desert, Chong *et al.* (1999) showed that, whereas both gypsum and anhydrite are stable in a Ca-Na-SO$_4$-H$_2$O system, the system only becomes oversaturated with respect to glauberite at Cl concentrations of about log Cl = log 0.5 molal. Once saturation with respect to glauberite is achieved, glauberite begins to precipitate at the expense of the dissolution of gypsum, leading to complete dissolution of the gypsum as salinity increases. The presence of these minerals therefore suggests that a saline, briny system was present within the channel.

The additional presence of saponite and hydrated silica can also be explained in terms of evaporative sequences in the Atacama. In a study performed on the nearby Salar de Atacama, Boschetti *et al.* (2006) showed that inflow water in equilibrium with Mg-montmorillonite would evolve into the stability field of saponite, and then migrate toward saturation of magnesite, amorphous silica, and sepiolite. The NIR spectrum of sepiolite shares the same absorptions as saponite in the 1900 to 2500 nm region. It is only distinguishable from saponite by the loss of contrast of the shorter-wavelength absorption of the 1400 nm doublet and a slight shift toward longer wavelengths (from 1404 to 1417 nm) of the longer-wavelength absorption. Hence, it is possible that sepiolite occurs with the saponite and would not have been readily identified. We see no evidence for the presence of magnesite in the NIR spectra nor in the XRD data.

Capping the glauberite horizon is a dominantly gypsum horizon of loose salt. This horizon is in turn capped by a set of anhydrite-rich plates whose exposed surface consists dominantly of gypsum.

Given the upward concavity of layers S9 (and perhaps S8) and upsection, we propose that the salts in the upper section of WIN ridge were likely precipitated within the already carved fluvial channel. Given that the composition of layers S11-S13 is consistent with a playa-like deposit, it is possible that the channel was very shallow (0.3 – 0.5 m) and was filled almost to the brim by evaporite deposits S10-S14.

*Capping plates:* The knobby morphology and composition of the surface of the capping plates is intriguing. Although the plates contain a mixture of anhydrite and gypsum, the exposed surface is gypsum. We suggest that the observed knobby texture results from repeated dissolution and re-precipitation of Ca-sulfate, which may occur on a nearly daily cycle. Although this portion of the Atacama Desert is exceptionally dry, the area is often fog-covered in the mornings. Bracconi *et al.* (2010) found that anhydrite will deliquesce at 97–98% relative humidity (RH) at 20°C and re-precipitate gypsum. Pure gypsum itself does not have a deliquescence point. However, gypsum is fairly soluble (2.14 g/l at 20°C), and in the presence of halite (which deliquesces at 75.4% relative humidity, 20°C), it will dissolve at 90% RH and only precipitate out once the RH has dropped below 75%. In this scenario, a thin layer of gypsum may be dissolved by the fog every morning, remobilizing and re-precipitating as the fog lifts. The knobby texture would result from nucleation during precipitation, and any incipient instability will then control the spatial distribution of the next episode of re-precipitation, leading to a self-reinforcing phenomenon. A similar explanation applies for the digitate protrusions observed at the edges of the channel, which also exhibit a knobby texture at mm scales (**Fig. 9**). The field team has been present during the morning fog, but has not directly observed wet conditions associated with the capping plates. Alternatively, the capping platesmay have formed in standing water given their somewhat similar morphology to gypsum precipitates in nearby perennial brine pools in the Salar de Llamara, located southwest from the study site. (Rasuk et al., 2014).

Preservation of the WIN ridge: Inverted channels form as a consequence of preferential erosion of the surrounding sediment relative to the channel deposits because the channel surface acts as a protective capping unit. Most inverted channels described in the terrestrial literature exhibit a surface that is either cemented by precipitates (*e.g.,* sulfates, carbonates, silica, etc.), covered by volcanic flows, or covered in clasts that cannot be transported by the wind (*e.g.,* Zaki et al., 2018). In the case of the WIN ridge, the top surface is covered with loose detritus and decimeter-sized sulfate plates that provides some protection from deflation, but the main reason the ridge survived erosion is due to two cementation mechanisms. Other than the top 7 cm (S14), which consist of loose salts, the ridge is fairly well indurated throughout its volume, presumably by anhydrite, which is prevalent throughout the section. In addition to the anhydrite, a slope-conforming armor composed dominantly of halite appears to have formed since the formation of the ridge, and it also protects the ridge from erosion. Whereas the cementation of channel sediments is often the main mechanisms for channel inversion reported in the literature, the two cementation mechanisms we document at WIN (internal cementation and slope armoring) have not been previously recognized. The conditions that lead to these two cementation processes may be relevant to the inversion of channels in arid, sulfate-rich environments, such as those found on Mars, and are worth exploring further.

*Sulfate cements*: The preponderance of anhydrite as the dominant form of calcium sulfate within the ridge is intriguing. Gypsum tends to be the predominant Ca-sulfate precipitate in shallow-depth evaporitic environments, with anhydrite becoming predominant toward depths of 450 m (Strakhov, 1962; Sonnefeld, 1984; Klimchouk, 1996) or temperatures greater than 42°C (Charola *et al.*, 2007). Although these temperatures can conceivably be achieved in the Atacama (modern temperatures reported by Chong *et al.* (1999) range between highs of 35°C in the summer to lows of about 10°C in the winter), the precipitation of anhydrite at these temperatures is kinetically inhibited (Freyer and Voigt, 2003). Voigt *et al.* (2019) showed that the Central Depression of the Atacama (where the WIN ridge is located) is a unique region of the Atacama in which anhydrite is dominant relative to gypsum. They cautioned, however, that although some correlation to aridity exists, the explanation is not straightforward, and they argued that the most likely scenario for the formation of anhydrite in the Central Depression of the Atacama was direct precipitation in a fluid with high salinity. Precipitation of β-anhydrite is thermodynamically favored over gypsum in environments with high chloride and nitrate concentrations (Hardie and Eugster, 1980; James, 1992; Charola *et al.*, 2007) due to the decreased thermodynamic activity of water. Sonnefelt (1984) showed that gypsum dehydration can occur at shallow depths during early diagenesis by interaction with hygroscopic brines of Na, Mg, or Ca chlorides; and Hardie (1967) showed that the thermodynamic stability boundary between gypsum and anhydrite could be as low as 18°C for saturated NaCl solutions. Given this finding, it would be reasonable to infer that the anhydrite throughout the WIN ridge was precipitated directly from solution, rather than forming from the dehydration of gypsum.

Sulfate cements are present throughout the section in abundances significantly greater than pore-filling. Precipitation of sulfate salts throughout the ridge may have occurred in an evaporative distal alluvial or playa-like setting, or post-deposition via early diagenesis. The distinction between these two environments can be inferred to some degree from the coloration and bedding of the sediment. Three important observations for the lower unit suggest that the sulfates were emplaced via early diagenesis as a consequence of precipitation out of groundwater:

1) White salt flakes decrease in concentration from the bottom of S2 to the top of S4. This upward decrease in flake concentration suggest a process of groundwater upwelling.
2) Mottling of layers S6-S8. Mottling within sedimentary units is thought to be due to seasonal groundwater, where reducing conditions and Fe mobilization occur during saturation, and oxidizing conditions leading to precipitation occur during drying.
3) Lack of sedimentary texture throughout the section. No bedding indicative of sedimentary processes (*e.g.,* fine layering, cross lamination, festoons) was observed, suggesting that any pre-existing sedimentary texture was erased by the precipitation and crystallization of sulfates.

*Slope-conforming armor*: The identification of a cm-thick slope-conforming armor (S8b-skin) composed dominantly of halite (about 50 wt%) and anhydrite (about 20 wt%) and containing minor quantities of other highly soluble minerals such as glauberite and polyhalite is particularly intriguing, as this armor must have formed after the topographic inversion took place. Remobilization of soluble salts in the Atacama is an ongoing geological process that acts on timescales of years to decades as halite, which is extremely soluble, is slowly wicked by moisture toward the surface of the landform.

Depositional plain: Aeolian deflation is the dominant geomorphic process in the hyperarid environment where the WIN ridge formed. Testament to this dominance is the extensive field of yardangs surrounding the ridge, composed of the sedimentary rock into which the WIN channel was carved. The grain size, bedding, and composition of the yardangs offer clues to its geological history. The basal well-sorted fine sand layers of yardang are capped by two sandy mud layers. The upper sandy mud likely records overbank deposits. The lower sand may be of eolian or fluvial origin. Root holes and orange stains were observed in the uppermost, fine-grained section, indicating a vegetation-supporting environment. Fine layering was not observed, which may be due to episodic deposition of thick layers of sediment, or destruction of the sedimentary fabric by precipitation of salts within the rock mass in early-stage diagenesis. No erect plant stalks were found in the stratigraphy, as can occur during rapid deposition of sediment in a flood-like setting, suggesting that sediment deposition may have occurred on barren surfaces and then groundwater rose to support vegetation in the capping unit.

Mineralogically, the yardang consists dominantly of a combination of primary minerals (consistent with a felsic source), calcium sulfates, and some halite. XRD indicates a fairly large abundance of calcium sulfates in the rock (50% on average with a variance of 20%), dominantly anhydrite in the lower portion of the section and transitioning to dominantly gypsum toward the upper portion. As previously discussed, anhydrite can form from the dissolution of gypsum and its re-precipitation in briny solutions. The observed enhancement in in halite concentration directly above the anhydrite-rich layer suggests that a saline environment may have been present during early diagenesis of the lower, sandy unit. In contrast, the sulfate mineralogy of the upper, fine-grained unit is dominated by gypsum, which is thermodynamically preferred in a less saline evaporitic environment.

<u>Relationship between yardang and WIN ridge</u>: One of the most notable results from the study is the recognition that the sediments that make up WIN ridge and the yardangs contain a large amount of Ca-sulfates, implying an arid early diagenetic environment in which sulfates were incorporated into the rock mass, and perhaps removing the sedimentary texture in the process. Both the ridge and the yardang exhibit orange stains likely associated with jarosite, resulting from the decomposition of organics in an anoxic environment. However, these stains are not necessarily indicative of groundwater flow. In contrast to the yardang, mottling in the layers directly beneath the channel floor (S6-S7) is interpreted as evidence for ground water saturation in those layers of the lower unit, suggesting that the channel provided a better pathway for the flow of groundwater than the surrounding terrain.

Given that the lower unit of WIN ridge has a mineralogy similar to that of the yardang, we propose that the WIN channel originally carved down to the mottled layers (S6 and S7; an incision depth of at least 40 cm). If the lower unit of WIN ridge was deposited in the same event(s) as the yardang, then similarities in the grain size, mineralogy, and occurrence of hardpans should be observed. A hardpan occurs at the base of both the yardang and the ridge. The mineralogy of this hardpan is fairly consistent between the two landforms. The bottom 40 cm in both landforms consist of sediments with sand-sized grains as well as white flakes, transitioning to finer grain sizes and a decrease in white flakes upward. However, the hardpan observed at 40 cm on the WIN ridge (S5) does not appear to occur at the yardang. From a compositional perspective, the lower section of the WIN ridge exhibits a primary mineralogy dominated by quartz, muscovite, and feldspar that is similar to that of the yardang (**Fig. 17**). The relative abundance of quartz is fairly consistent for both formations. However, the relative abundance of feldspar is notably greater in the yardang ridge. Unless the difference in these abundances can be accounted for by mineralogical variations of the amorphous component in both units, then there appears to be an intrinsic difference in abundances of feldspar and muscovite that cannot be explained by a co-deposition model. There are two potential answers for this compositional variation:

1) Potentially, there is lateral variability within the depositional system that explains the difference in primary mineral abundance between the yardang and lower WIN.
2) The chemical variability may not be a real difference, but might be related to the measurement technique. Here, we draw upon results from the Chemistry and Mineralogy (CheMin) X-ray diffraction instrument aboard the Mars Science Laboratory *Curiosity* rover because this instrument is functionally identical to Terra (Blake et al., 2012. CheMin analysis of two samples from the same rock (<1 m apart) had mineralogical variability interpreted to be reflective of grain motion in the sample holder during sample analysis rather than intrinsic differences in the rock (Rampe et al., 2019).

Geological history of WIN channel: The numerous layers identified in the WIN ridge indicate a complex history of deposition, erosion, and diagenesis. Here we attempt to reconstruct the sequence of events that led to the formation of this landform (**Fig. 18**). The lower section of WIN channel as well as the yardang indicate a period of intermittent deposition of primary material transported fluvially (S2-S4, S6), likely from the Andes Mountains to the east (**Figs. 18A-C**). Root holes and orange staining interpreted as remnants of vegetation are observed in these sediments, indicating that water was at least periodically available in the sub-surface. Anhydrite paleosols during periods between depositional events (e.g., S1, S5). Upwelling groundwater, either intermittently or continuously throughout this process, led to the precipitation of anhydrite cements throughout the subsurface. Subsequent or parallel to part of the depositional period, the WIN channel was carved by flowing water, which was accompanied by water saturating the materials beneath the channel floor (**Fig. 18D**), resulting in the mottling of these materials (S7), as well as deposition of sediments (S8-S9, **Fig. 18E**). The root tracks in S9 suggest that these deposits were exposed subaerially for long enough, and that the salinity was low enough, for vegetation to grow (the low salinity may have also allowed the precipitation and preservation of gypsum, rather than anhydrite or glauberite). Subsequently, the surface was exposed to a playa-like environment with salinity increasing over time resulting in the formation of glauberite, halite, saponite, and hydrated silica (S10-S13). These minerals were eventually capped by a gypsic/anhydritic paleosol (**Fig. 18F**). Subsequent erosion of the surrounding sediment resulted in the presently observed terrain morphology. Finally, precipitation of a halite crust on the side slopes armored the ridge (**Fig. 18G**).

**Conclusions:**
We performed a stratigraphic and mineralogical analysis of a vertical transect across a ridge interpreted to be an inverted channel. This inverted channel is part of a system of alluvial deposits in the Pampa del Tamarugal of the Atacama Desert, Chile. We found a history dominated by diagenesis that includes groundwater mobilization and precipitation of anhydrite cements throughout the volume of the ridge. The section can best be described as a sequence of horizons including, from the bottom up: an anhydrite paleosol at the base (S1), an anhydrite-cemented sandstone consisting primarily of mostly unaltered alluvial detritus (S2-S4), a second anhydrite paleosol (S5), an anhydrite-cemented and mottled jarosite-rich horizon (S6-S8), a layer consisting of gypsum-rich material (S9), an anhydrite paleosol (S10), an anhydrite-rich glauberite-bearing playa-like deposit (S11-S13), and a fourth paleosol consisting of gypsum and anhydrite (S14).

We also identified two mechanisms that contribute to channel inversion and preservation. The first mechanism is the cementation of the materials of the lower unit by anhydrite cements, which are interpreted to have precipitated from upwelling groundwater. Sulfate cements were also observed in erosional remnants of the surrounding terrain, although not at the same abundance as within the mass of the inverted channel. To our knowledge, this is the first documented case of anhydrite as the dominant induration agent in an inverted channel. The second mechanism that we identified is the armoring of the lateral slopes of the ridge by a halite-rich cement. The slope-conforming armor formed by this second mechanism appears to have developed subsequent to the formation of the ridge as a consequence of the remobilization of soluble salts. These newly identified mechanisms should inform us towards a better understanding of the processes and environmental conditions involved in channel inversion on Mars.

**Acknowledgments:** This research was supported by grants to R.M.E. Williams from the NASA Mars Fundamental Research Program (NNX13AG83G) and the NASA Mars Data Analysis Program (80NSSC19K1216).

**References:**

Blake, D. et al., (2012). Characterization and calibration of the CheMin mineralogical instrument on Mars Science Laboratory. *Space Science Reviews*, 170(1-4), 341-399, DOI:10.1007/s11214-012-9905-1.

Bracconi, P., Cyrille, A., and Mutin, J.-C. (2010) Interaction of CaSO4 with water vapour at high relative humidity, Open archive (HAL), available at: https://hal.archives-ouvertes.fr/hal-00454539.

Buurman, P. (1975) Possibilities of palaeopedology. *Sedimentology* 22, pp 289-298

Burr, D. M., R. M. E. Williams, K. D. Wendell, M. Chojnacki, and J. P. Emery (2010), Inverted fluvial features in theAeolis/Zephyria Plana region, Mars: Formation mechanism and initial paleodischarge estimates, *J. Geophys. Res.,115,* E07011,doi:10.1029/2009JE003496.

Carlson, R, W. Smythe, K. Baines, E. Barbinas, R. Burns, S. Calcutt, W. Calvin, R. Clark, G. Danielson, A. Davies, P. Drossart, T. Encrenaz, F. Fanale, J. Granahan, G. Hansen, P. Hererra, C. Hibbitts, J. Hui, P. Irwin, T. Johnson, L. Kamp, H. Kieffer, F. Leader, R. Lopes- Gautier, D. Matson, T. McCord, R. Mehlman, A. Ocampo, G. Orton, M. Roos-Serote, M. Segura, J. Shirley, L. Soderblom, A. Stevenson, F. Taylor, A. Weir, P. Weissman (1996) Near-Infrared Spectroscopy and Spectral Mapping of Jupiter and the Galilean Satellites: First Results from Galileo's Initial Orbit. Science, 274, 385-388.

Charola, A.E., J. Pühringer, and M. Steiger, (2007) Gypsum: A review of its role in the deterioration of building materials. Environ. Geol. 52, 2, 339-352. doi: /10.1007/s00254-006-0566-9.

Chong, G, P.L. López, F. Luis L.F. Auqué, and I. Garcés (1999) Características geoquímicas y pautas de evolución de las salmueras superficiales del Salar de Llamara, Chile. Revista Geológica de Chile. Vol. 26. No. 1. p. 89–108.

Clark, R. N., R. O. Green, G. A. Swayze, G. Meeker, S. Sutley, T. M. Hoefen, K. E. Livo, G. Plumlee, B. Pavri, C. Sarture, S. Wilson, P. Hageman, P. Lamothe, J. S. Vance, J. Boardman I. Brownfield, C. Gent, L. C. Morath, J. Taggart, P. M. Theodorakos, and M. Adams (2001) Environmental Studies of the World Trade Center area after the September 11, 2001 attack. U. S. Geological Survey, Open File Report OFR-01-0429, (approximately 260 pages printed). http://pubs.usgs.gov/of/2001/ofr-01-0429/

Clark, R.N., G.A. Swayze, K.E. Livo, R.F. Kokaly, S.J. Sutley, J.B. Dalton, R.R. McDougal, and C.A. Gent, (2003). Imaging spectroscopy: Earth and planetary remote sensing with the USGS Tetracorder and expert systems, Journal of Geophysical Research, Vol. 108(E12), 5131, doi:10.1029/2002JE001847, p. 5-1 to 5-44. http://speclab.cr.usgs.gov/PAPERS/tetracorder

Clark, R.N., G.A. Swayze, T.M. Hoefen, R.O. Green, K.E. Livo, G., Meeker, S. Sutley, G. Plumlee, B. Pavri, C. Sarture, J. Boardman, I., Brownfield, and L.C. Morath (2006) Chapter 4: Environmental mapping of the World Trade Center area with imaging spectroscopy after the September 11, 2001 attack: in *Urban Aerosols and Their Impacts: Lessons Learned from the*
14

Table 1. Layer characteristics

| Layer | Interpretive name | Base Height (cm) | Thickness (cm) | Mineralogy[1] XRD (Major / Minor) | VNIR | Notes |
|---|---|---|---|---|---|---|
| Plate Top | **Reticulated plate (top)** | 103 | <1 | Anh / Bas, Gyp, Qtz, Plag, Mus | | Top surface of capping plates |
| Plate Bottom | **Reticulated plate (bottom)** | 103 | <1 | Anh / Bas, Qtz, Plag | | Bottom surface of capping plates |
| S15 | **Gypsum/Anhdrite surface** | 103 | <1 | Gyp, Anh / Qtz | Gyp | Poorly cemented material. between and under capping plates |
| S14 | **Gypsum Horizon** | 96 | 7 | Gyp, Anhe / Qtz | Gyp | Fluff of loose salt - white. Capped by sulfate plates. |
| S13 | **Glauberite+Halite Horizon 3** | 88 | 8 | Anh, Glb / Qtz, Hlt | Sap, Opl | Fe stain blob some black stain. |
| S12 | **Glauberite+Halite Horizon 2** | 88 | 8 | Anh / Qtz, Hlt, Glb | Sap, Opl | Lightly cemented, horizontally layered slabs of salt rich deposit. |
| S11 | **Black Glauberite & Halite Horizon 1** | 85 | 3 | Anh / Qtz, Hlt, Glb | Mont | Blackish sands. |
| S10 | **Anhydrite Horizon 3** | 77 | 8 | Anh / Qtz | Mont | Friable granular soft salt, white. |
| S9 | **Gypsum Horizon** | 74 | 4 | Gypsum, Qtz, Mus / Anh, Plag | Mont | Fine sands (with mica), clay, large Fe stained root casts. |
| S8b | **Halite Armor** | N/A | N/A | Hlt, Anh / Qtz, Mus, Glb | | Strong cm-thick slope conforming armor. |
| S8 | **Orange Jarosite Horizon** | 71 | 2 | Anh / Jar, Qtz | Jar, Bas | Fe-stained sands, mottled, white flakes. |
| S7 | **Mottled Jarosite Horizon** | 50 | 22 | Anh, Qtz, Jar / Plag, Mus | Mont | Tan sandstone with red, purple, orange, yellow mottling and white flaking. Bioturbated mud with yellow blotches. |
| S6 | **Purple Jarosite Horizon** | 41 | 9 | Anh, Qtz, Plag, Jar / Mus | Mont | Dark-toned purple, well-sorted sandstone with red stains and root holes. |
| S5 | **Anhydrite Horizon 2** | 41 | 1 | Anh / Hlt, Qtz, Plag, Mus | Mont | Thin (1 cm), hard salt layer. |
| S3, S4 | **Primaries Layer 2** | 16 | 25 | Anh, Qtz / Plag, Mus | Mont, Ill, Pxn | Fine sandstone with white salt flakes. Records vertical decrease of white mottling from base upward. Contains orange root casts. Top surface exhibits a scalloped depression (6 cm circular scallops). Contains secondary film of mineralization (S5) on top of layer. |
| S2 | **Primaries Layer 1** | 1 | 15 | Anh, Qtz, Mus / Plag | Mont, Ill, Pxn | Fine sandstone layer. Contains white and orange mottling. |
| S1 | **Anhydrite Horizon 1** | 0 | 1 | Anh / Qtz, Plag, Mus | Mont | Tan hardpan at ground level. Rough texture. |
| Y4 | **Gypsum Horizon 3** | 61 | 21 | Plag, Qtz, Gyp / Anh, Mus | Mont | Mud with some sand; some Fe-stain; root holes. |
| Y3 | **Gypsum Horizon 2** | 34 | 27 | Plag, Gyp / Anh, Qtz, Mus, Hlt | Mont | Mud with some sand; some Fe-stain. |
| Y2b | **Gypsum Horizon 1** | 20 | 14 | Qtz, Plag, Gyp / Anh, Mus, Hlt | Mont | Well sorted fine sand with speckles of salt. |
| Y2 | **Anhydrite Horizon 3** | 18 | 2 | Anh, Qtz, Plag, Gyp / Mus | Mont | Orange stain. |
| Y1 | **Anhydrite horizon 2** | 0 | 18 | Anh, Qtz, Plag, Hlt / Mus | Mont | Sand with salt; well sorted, fine grained salt. |
| Y1b | **Anhydrite horizon 1** | -1 | 1 | Anh / Qtz, Mus, Plag | Mont | Sandy unit with salt. |

[1]Sap = saponite; Anh = anhydrite; Bas = bassanite; Gyp = gypsum; Qtz = Qtz; Plagioclase = Plag; Mus = muscovite; Hlt = halite; Glb = glauberite; Jar = jarosite; Ill = illite; Pxn = pyroxene; Mont = montmorillonite; Opl = opal

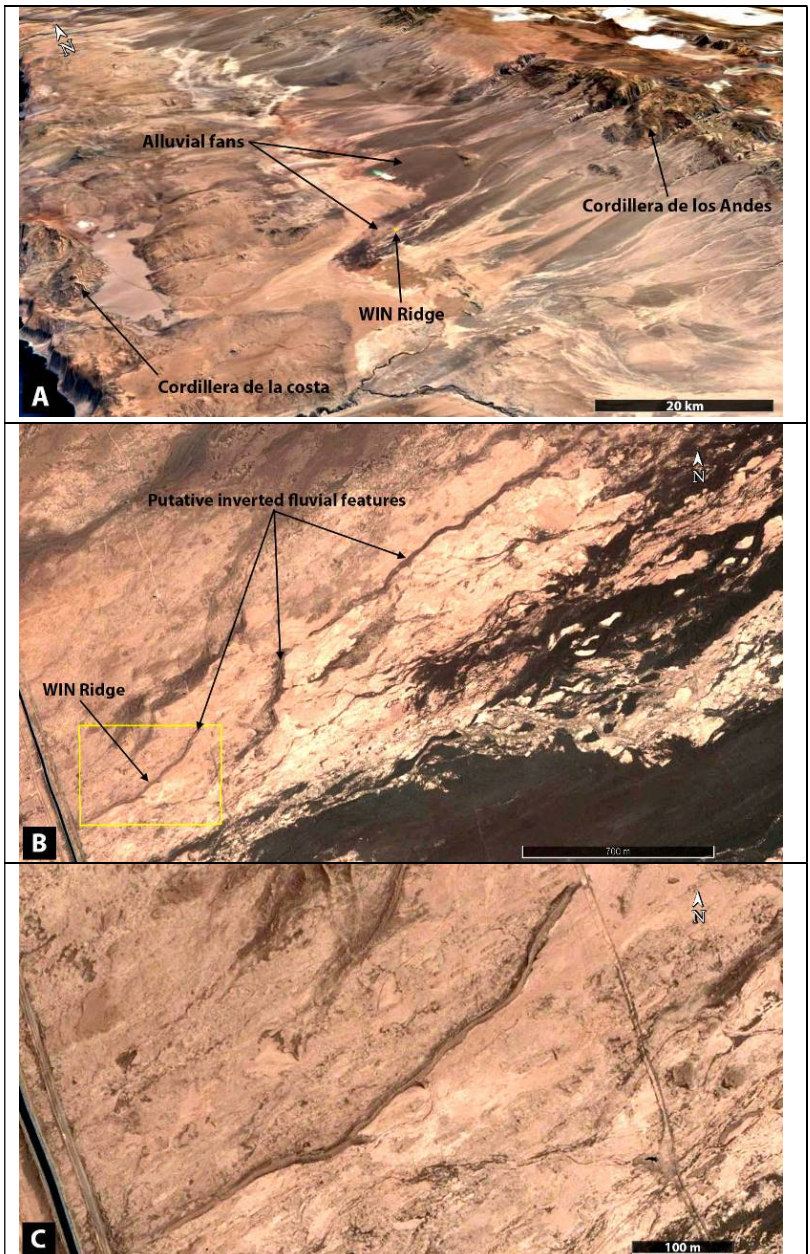

**Figure 1.** (A) Perspective view of the Pampa del Tamarugal region, showing the western Andes (right) Cordillera de la Costa (left), and the highly dissected alluvial fans. WIN Ridge is located at the yellow spot indicated by the arrow. Image rendered in Google Earth using Landsat imaging overdrapped on topography at 3x vertical exaggeration. (B) Plan view of the distal end of a fan sourced from the western Andes. Multiple ridges interpreted to be inverted channel segments are apparent in the region. WIN ridge is show in the inset. (C) Plan view of WIN ridge. Images produced for Google Earth by Maxar Technologies.

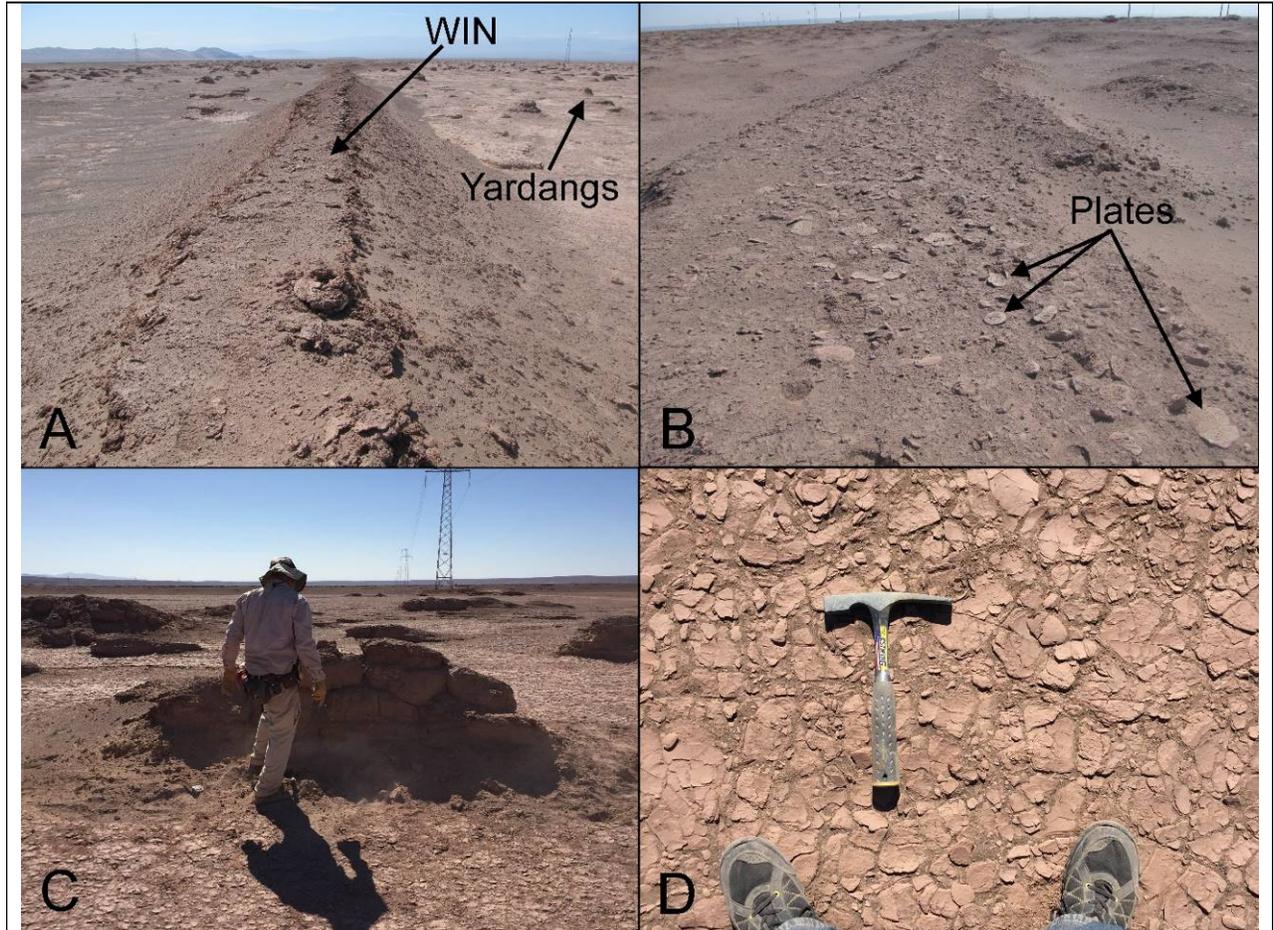

**Figure 2.** (A) View along the top of the WIN ridge, with the yardang field in the background to the left (B) Another section of WIN ridge showing capping plates (Fig. 8A). (C) Yardang studied in this work. D) Texture of inter-yardang terrain shows decimeter-scale polygons.

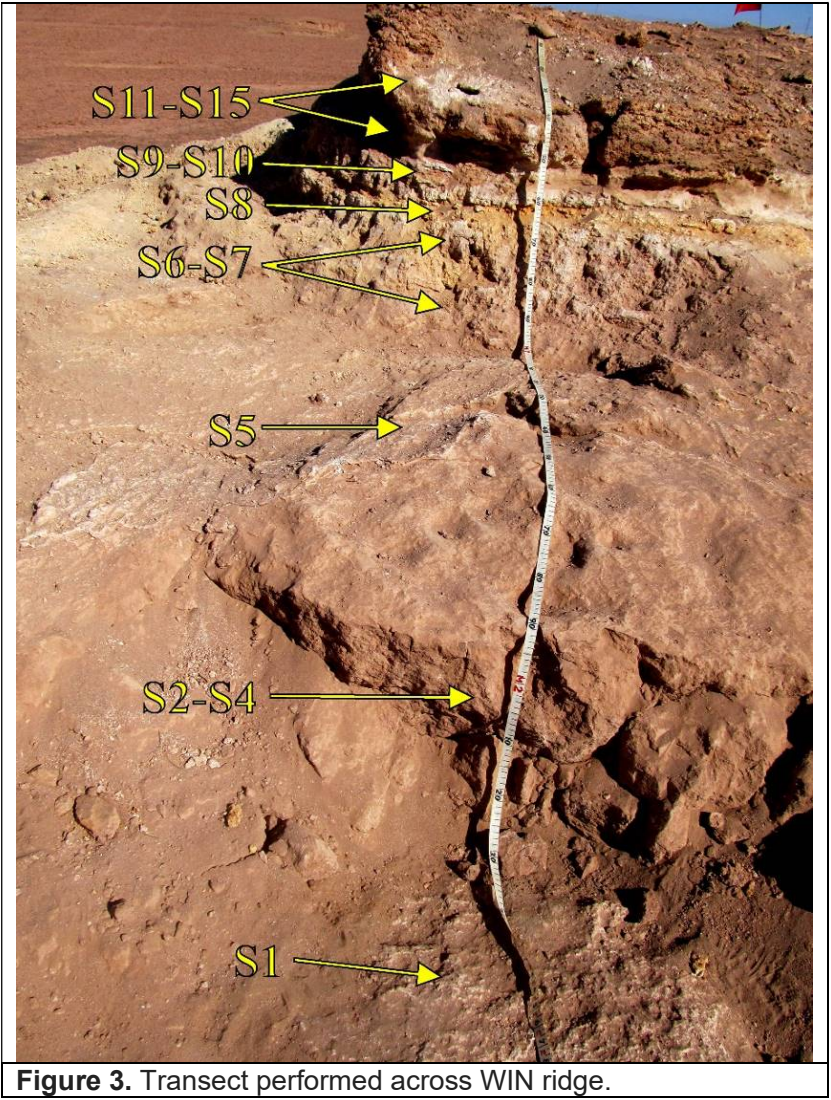

**Figure 3.** Transect performed across WIN ridge.

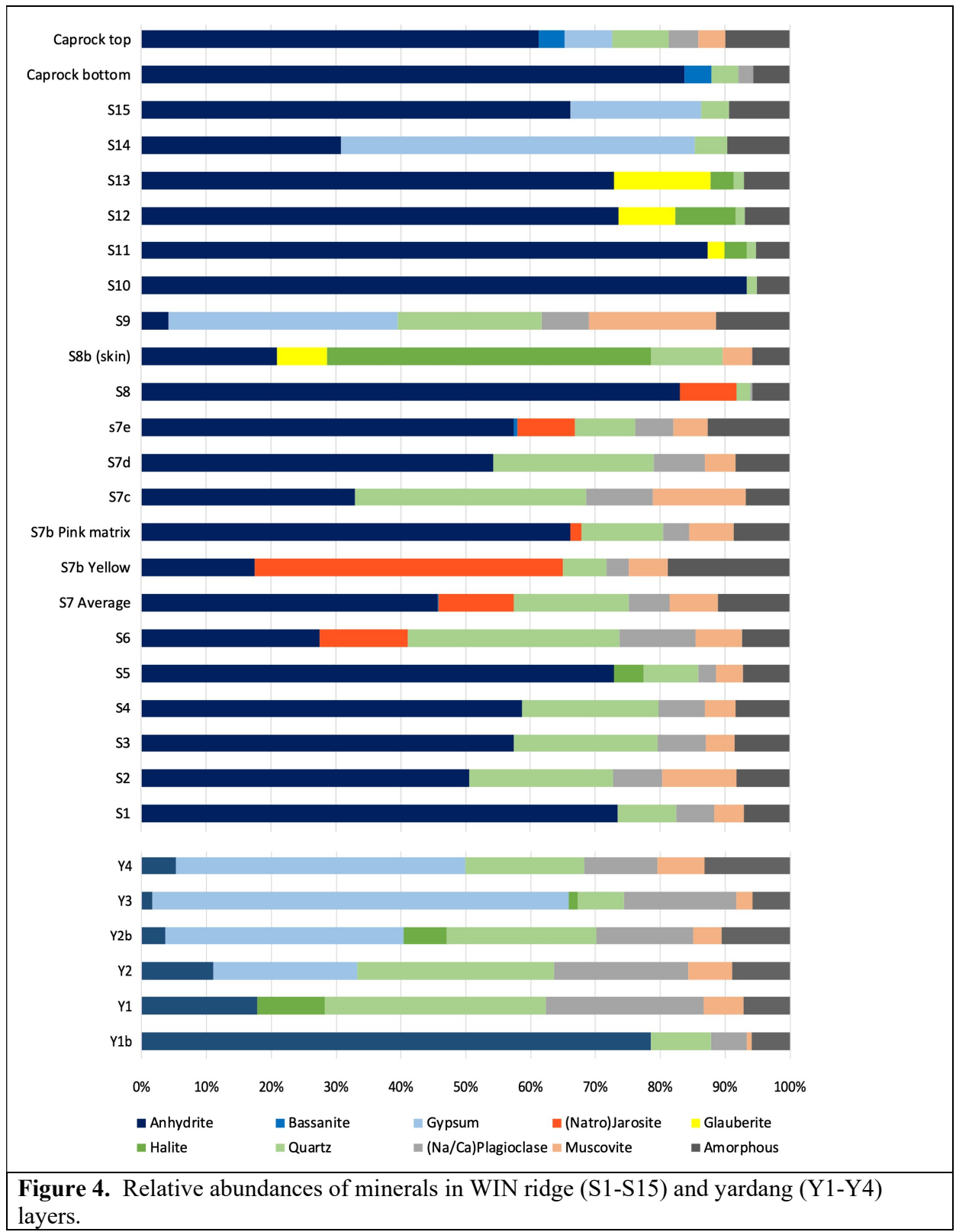

**Figure 4.** Relative abundances of minerals in WIN ridge (S1-S15) and yardang (Y1-Y4) layers.

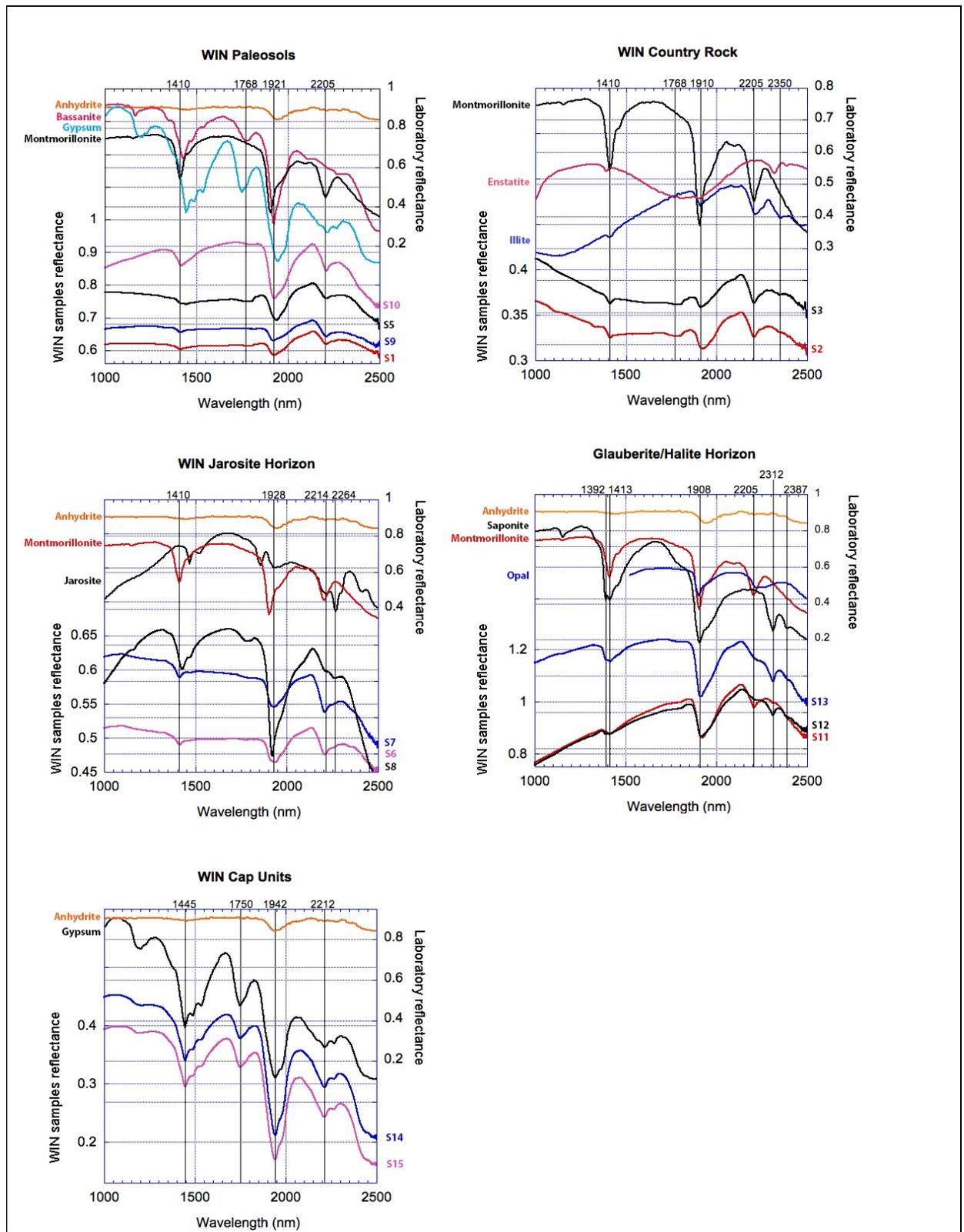

**Figure 5.** NIR spectra of WIN and yardang compared to laboratory spectra of candidate minerals. Plots are grouped by horizons as identified from the XRD analysis. Laboratory spectra are shown on top of each plot, and sample spectra on the bottom.

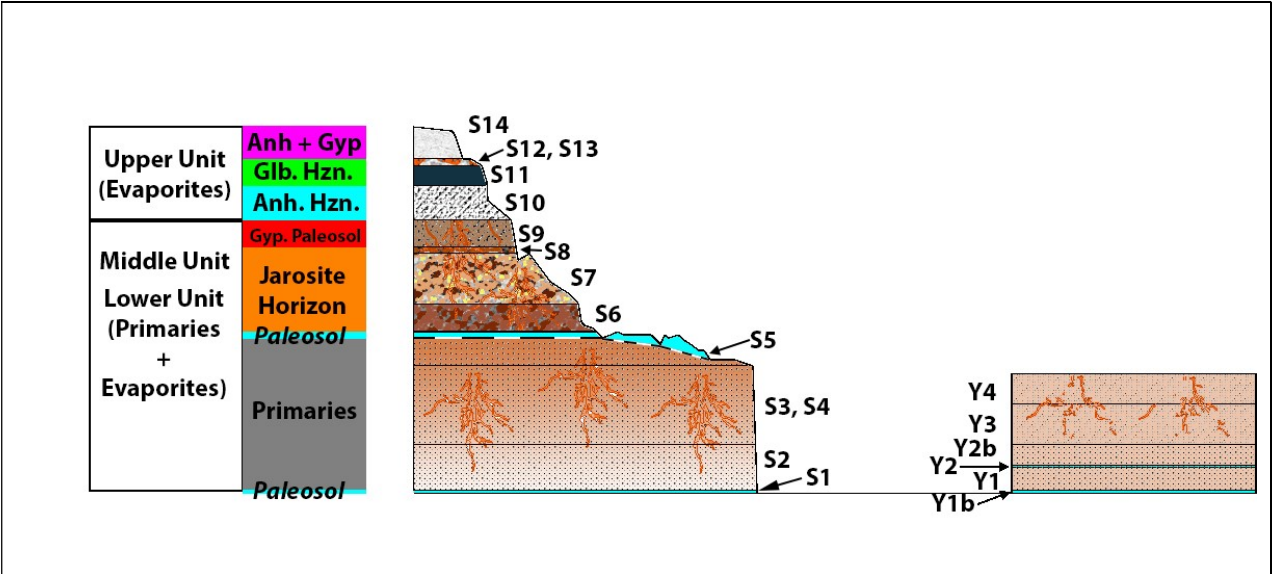

**Figure 6.** Stratigraphic cross section of WIN ridge (left) and nearby yardang (right). The outline of the ridge cross section was obtained from differential GPS measurements.

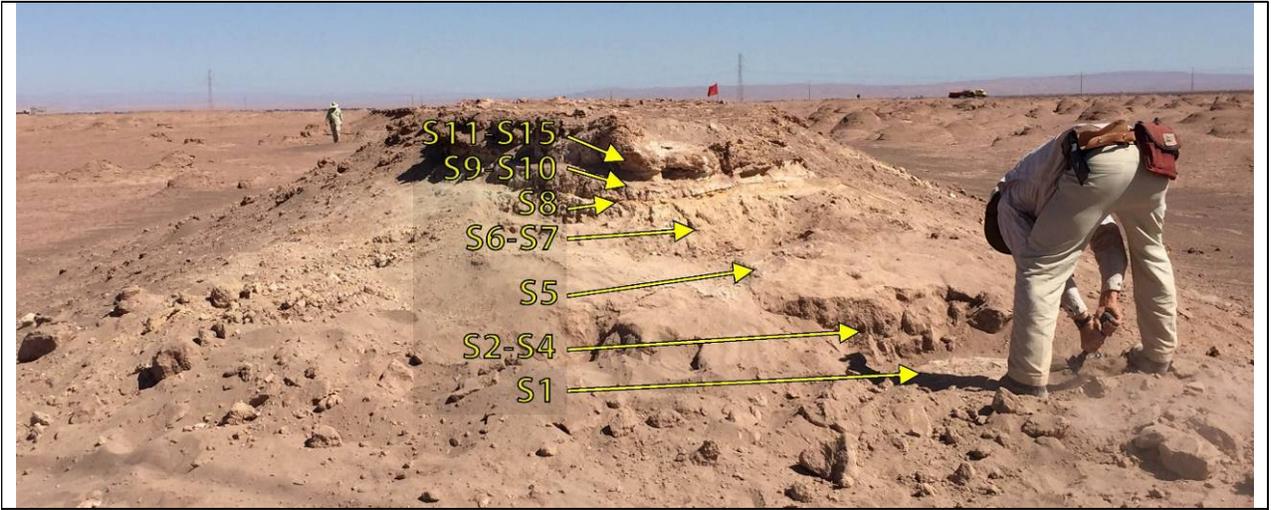

**Figure 7.** Frontal view of WIN ridge cross-section showing concave-up layering.

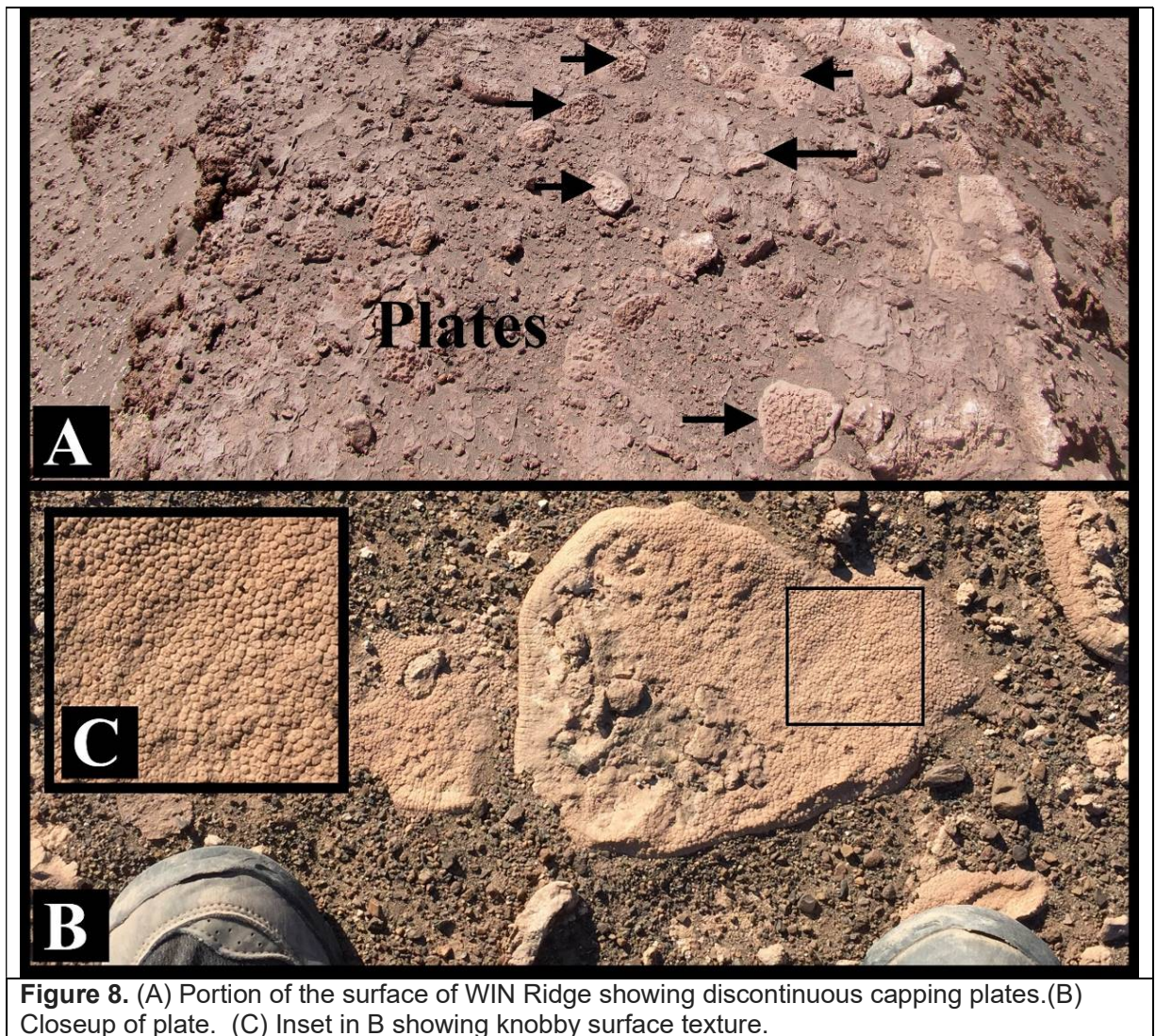

**Figure 8.** (A) Portion of the surface of WIN Ridge showing discontinuous capping plates. (B) Closeup of plate. (C) Inset in B showing knobby surface texture.

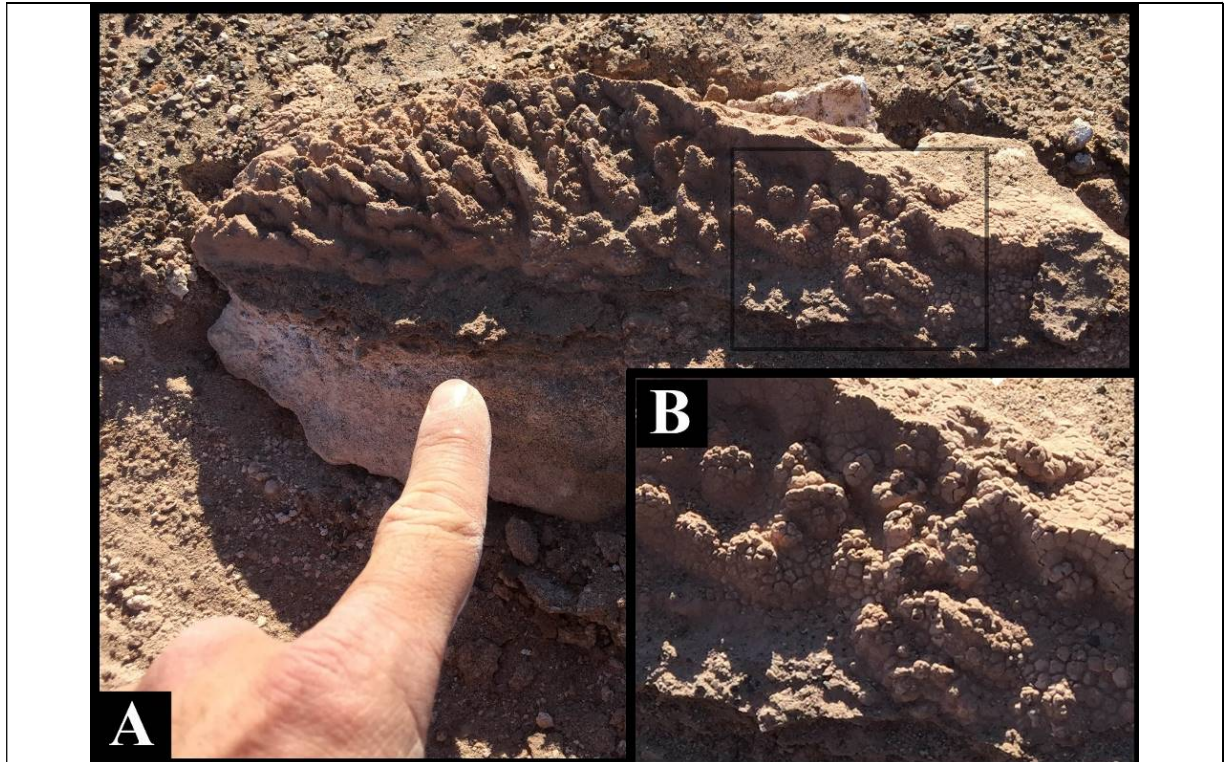

**Figure 9.** (A) Transition of knobby platy surface to digitate morphology at edges of the capping unit at WIN ridge. (B) Enlargement of black box in A. The knobby texture is still apparent on the right side of the rock.

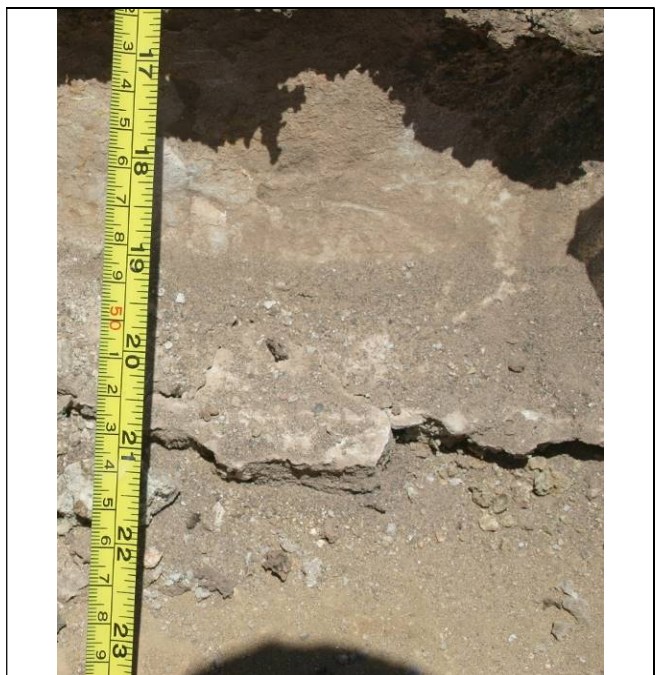

**Figure 10.** Surface-conforming anhydrite layer. This unit is highly indurated and is found draping the side escarpments of the ridge. Top is up. Shadows of digitate armor is seen at the top.

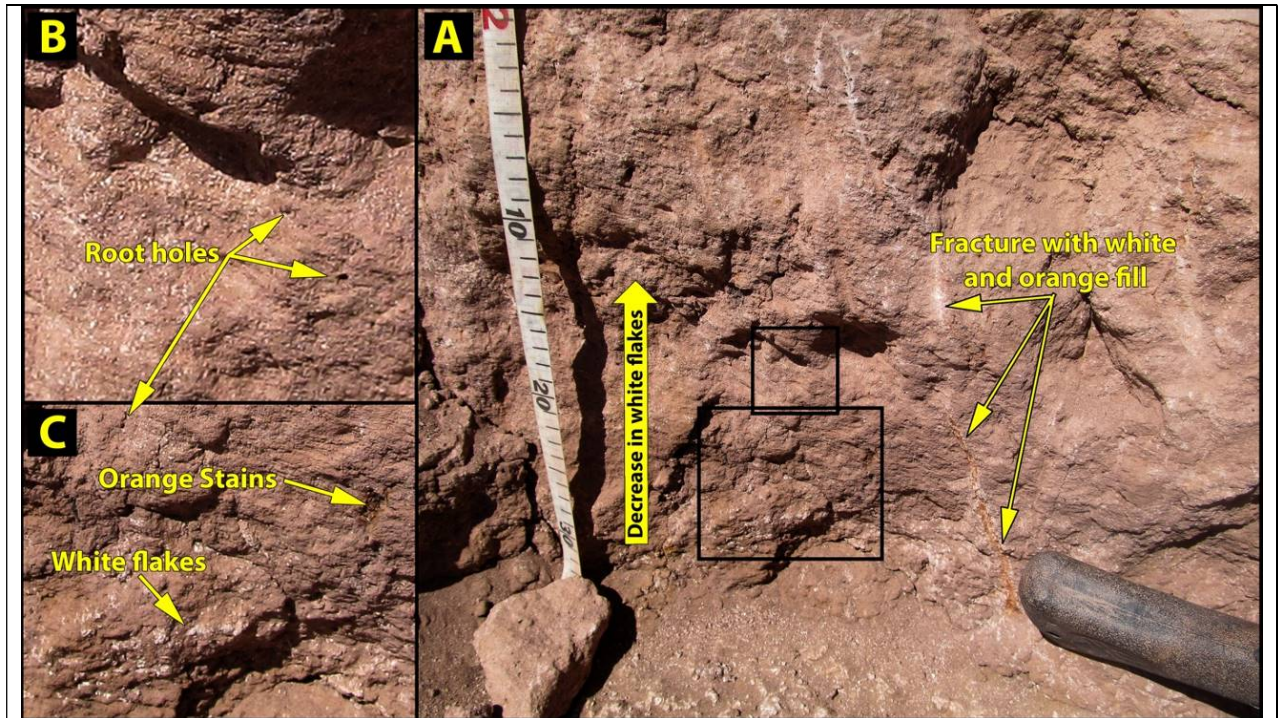
**Figure 11.** Bottom portion (S1-S4) of the lower unit. Smaller inset corresponds to B. Large inset corresponds to C.

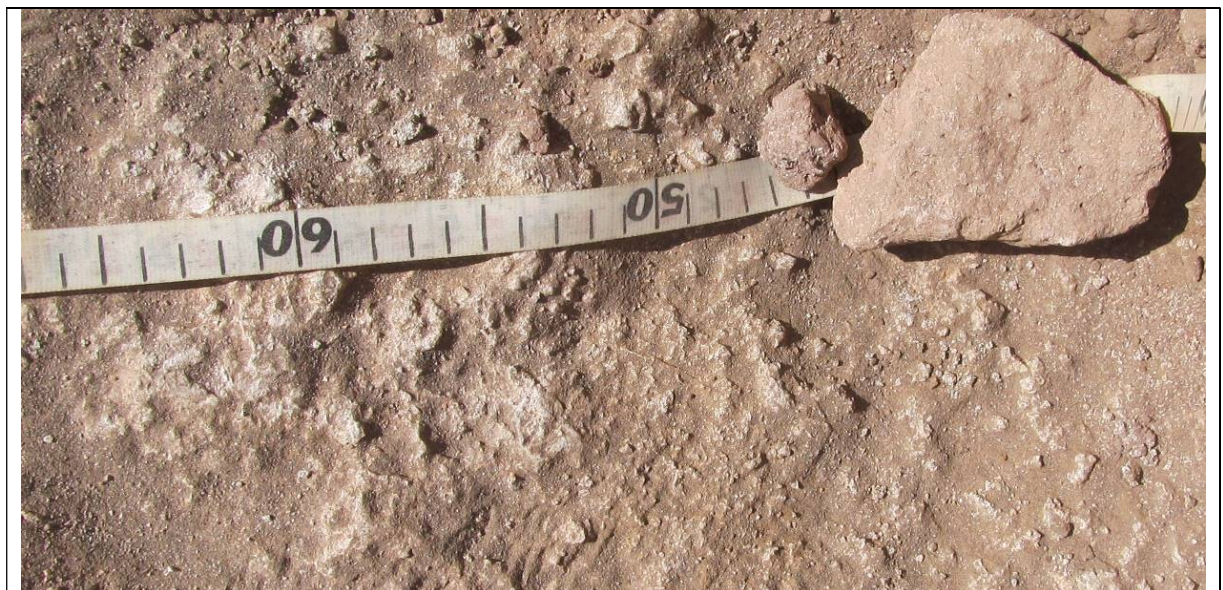
**Figure 12.** Surface of S1 showing cm-scale surface roughness.

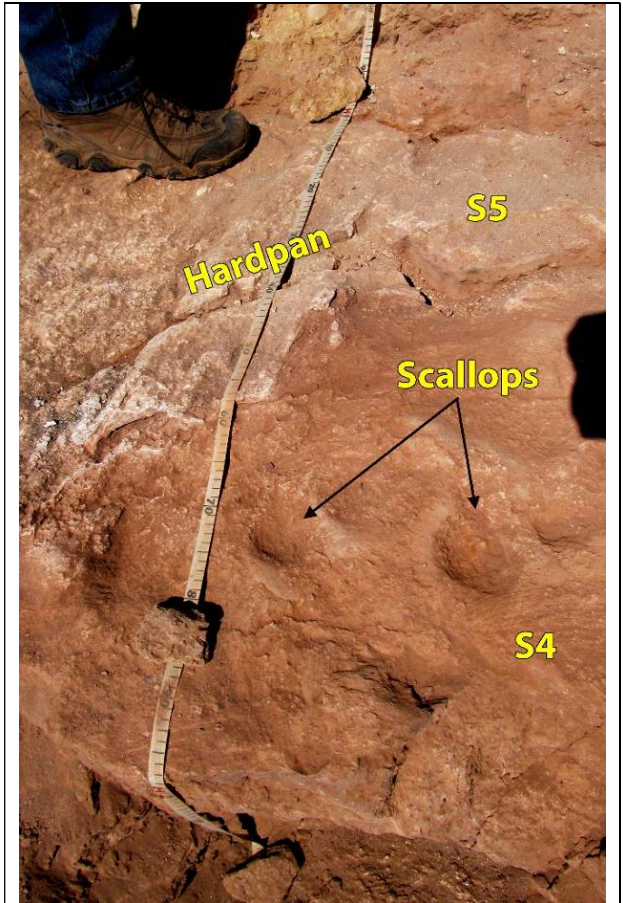

**Figure 13.** Surface of S4 showing scallps and surface mineralization (S5)

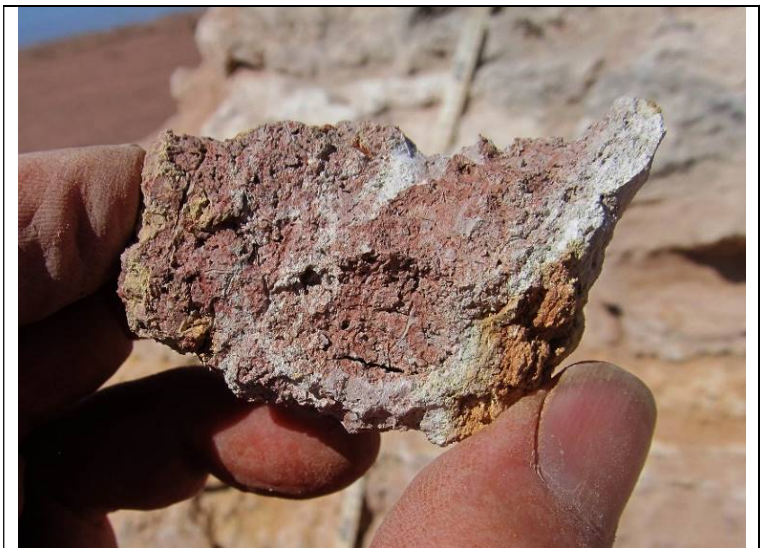

**Figure 14.** Materials within layer S7 exhibited significant heterogeneity in both coloration and composition.

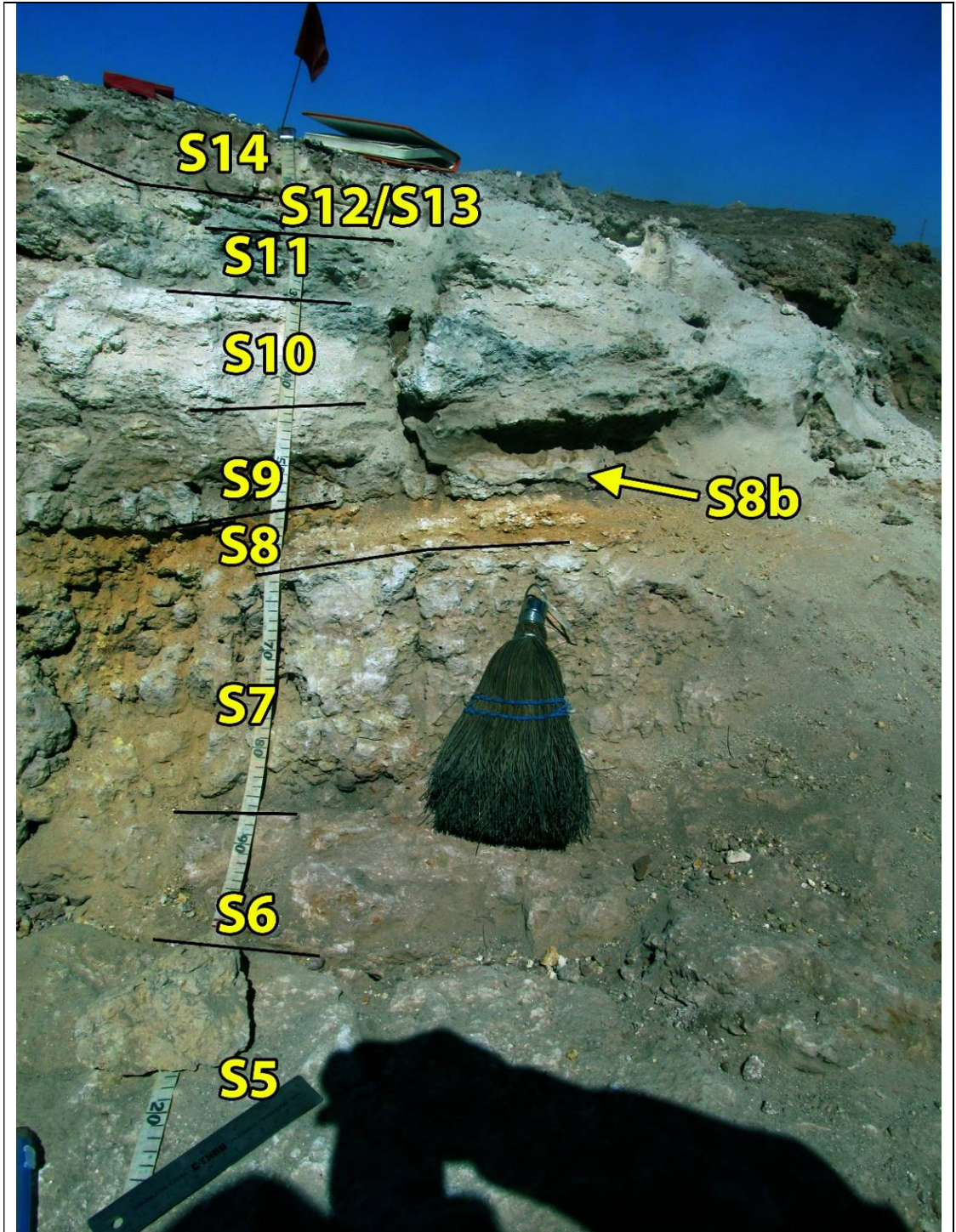

Figure 15. Layers S5-S14 showing color and texture variations throughout the section.

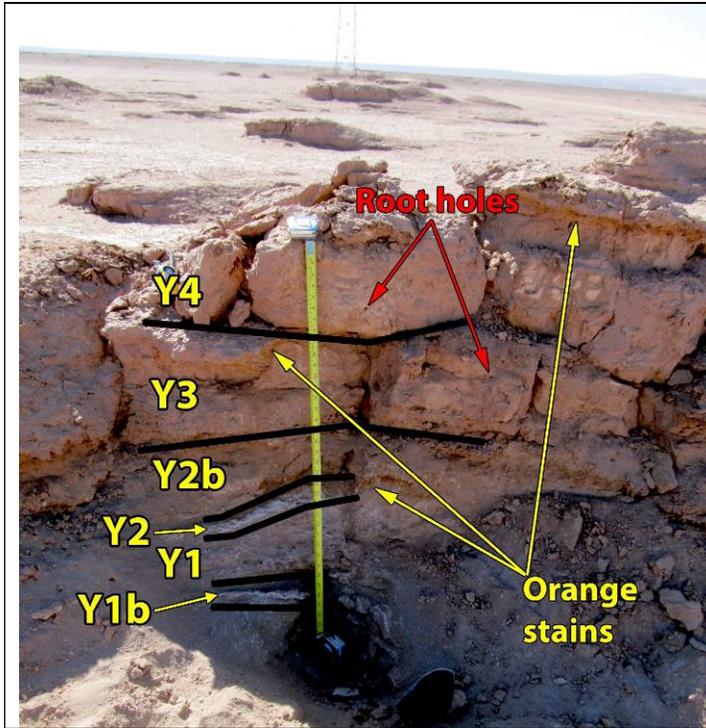

**Figure 16.** Yardang located about 14 meters from the WIN ridge transect. The yardang is approximately 82 cm tall and exhibits a sequence of thick internal layers.

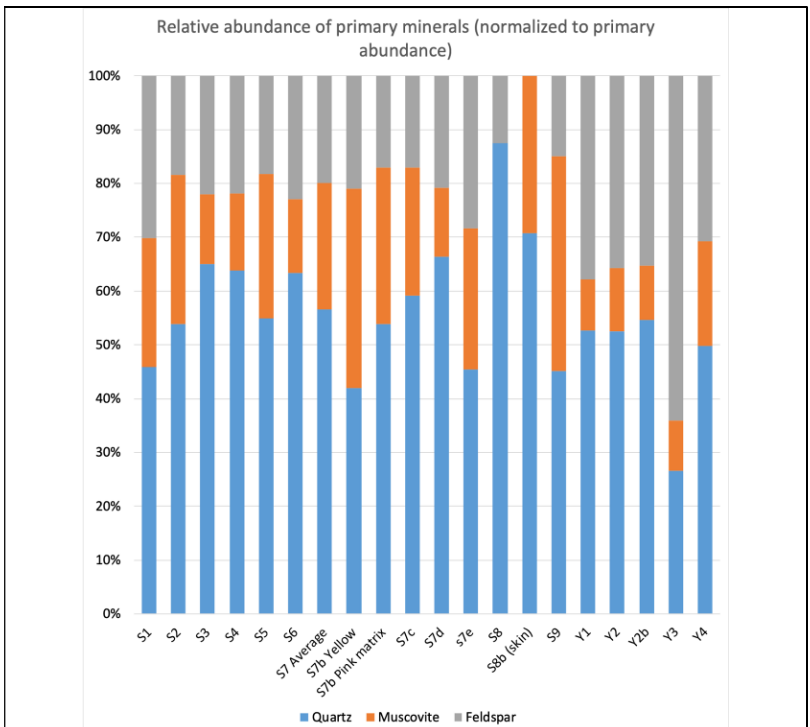

**Figure 17.** Relative abundance of primary minerals normalized to the sum of the primary mineral abundances.

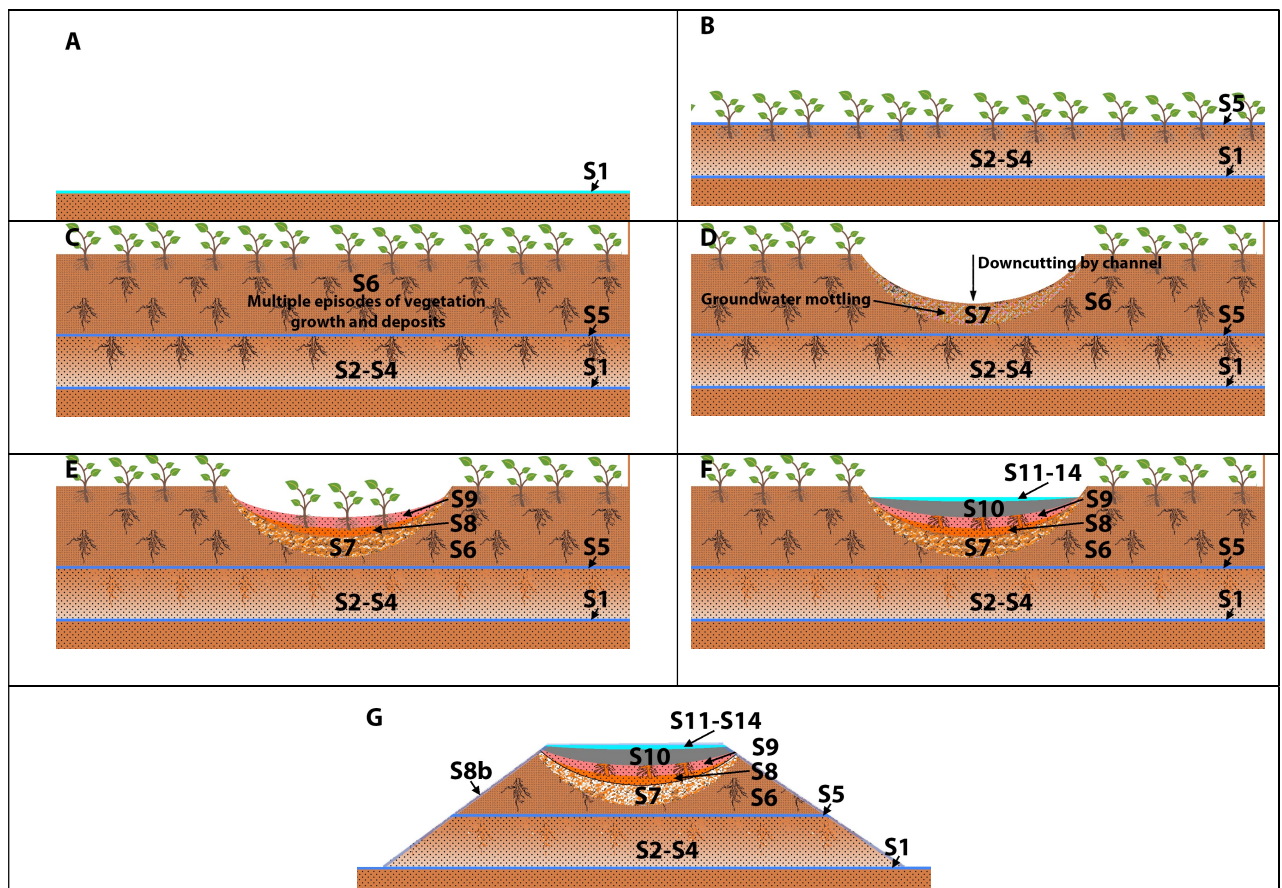

**Figure 18.** Inferred sequence of WIN ridge formation. A) Alluvial material (tan) capped with anhydrite paleosol S1 (blue). B) Additional deposition of primary minerals and vegetation growth (S2-S4) followed by the capping with anhydrite paleosol S5 (blue). C) $Fe^{3+}$ Staining of S2-S4 from decomposition of plant roots, sulfate precipitation from upwelling groundwater (white gradation) and burial by additional primary mineral deposits (S6-S7). D) Downcutting by WIN channel (black arrow) into sediments concurrent with groundwater flow (purple). E) Fluvial deposition of primary minerals (S8-S9) and jarosite precursor (S8), and formation of gypsic/anhydrite paleosol (S9-S10). Vegetation growth over S9. Layers (S6-S7) modified by groundwater flow appear mottled. F) Deposition of salt precipitates (S10-14) including glauberite/halite (gray - S11-S13) and gypsic paleosol (light blue – S14) and staining of S9 from decomposition of roots. G) Erosion to modern form and formation of armor (green - S8b).